\documentclass[prl,aps,twocolumn,groupedaddress,floats,showpacs,final,superscriptaddress]{revtex4-1}
\usepackage{dcolumn}
\usepackage{bm}
\usepackage{color}
\usepackage{mathtools}
\usepackage{amsmath,amsfonts,amssymb,bm,ulem}

\usepackage{graphicx}
\usepackage{epsfig}
\usepackage{psfrag}
\usepackage[pdftex]{hyperref} 

\begin{document}

\title{Dielectric function and thermodynamic properties of jellium in the GW approximation. }

\author{Kris Van Houcke}
\affiliation{Laboratoire de Physique Statistique, Ecole Normale Sup\'erieure, UPMC, Universit\'e Paris Diderot, CNRS, 24 rue Lhomond, 75231 Paris Cedex 5, France}
\author{Igor S. Tupitsyn}
\affiliation{Department of Physics, University of Massachusetts, Amherst, MA 01003, USA}
\affiliation{National Research Center Kurchatov Institute, 123182 Moscow, Russia}
\author{Andrey S. Mishchenko}
\affiliation{RIKEN Center for Emergent Matter Science (CEMS), 2-1 Hirosawa, Wako, Saitama, 351-0198, Japan}
\affiliation{National Research Center Kurchatov Institute, 123182 Moscow, Russia}
\author{Nikolay V. Prokof'ev}
\affiliation{Department of Physics, University of Massachusetts, Amherst, MA 01003, USA}
\affiliation{National Research Center Kurchatov Institute, 123182 Moscow, Russia}
\affiliation{Department of Theoretical Physics, The Royal Institute of Technology, Stockholm SE-10691 Sweden}

\date{\today}

\begin{abstract}
The fully self-consistent GW approximation is an established method for
electronic structure calculations. Its most serious deficiency is known to be
an incorrect prediction of the dielectric response. In this work we examine
the GW approximation for the homogeneous electron gas and find that problems with
the dielectric response are solved by enforcing the particle-number conservation
law in the polarization function. 
Previously reported data for the ground-state energy were plainly contradicting each other well
outside of reported error bounds. Some of these results created a false impression of how accurate the fully self-consistent GW approximation is.
We resolve this controversy by confirming that only Ref.~\cite{Yan} was reporting
correct energy data, and present values for other key Fermi-liquid properties.
\end{abstract}


\maketitle

Accurately solving the many-electron Schr\"odinger equation for real solid-state systems is a major challenge of great technological importance.
Among available theoretical approaches, approximations based on diagrammatic many-body perturbation theory\cite{Fetter,Mahan} are attractive 
because even at low order these approximations can grasp the essential physics and allow to deal with the long-range Coulomb interaction in the thermodynamic limit.
The widely used random-phase approximation (RPA), for example, can qualitatively
explain features of real metals such as screening, plasmon and Friedel oscillations.
In principle, the skeleton diagrammatic expansion allows one to systematically improve
on these results and obtain accurate solutions to the many-electron problem.
In practice, however, progress is hindered because more sophisticated lowest-order
diagrammatic approximations can lead to worse results, while a systematic evaluation
of the skeleton series seems computationally too expensive within the conventional implementation
(apart from questions about series convergence).

The most widely used diagrammatic method for electronic structure calculations
is the so-called GW approximation \cite{Hedin,Aryasetiawan,Onida}.
While clearly going beyond regular RPA by evaluating ``bubble''-diagrams in a self-consistent way,
the GW approximation has an additional advantage of being a conserving approximation
(with respect to the relation between the particle density $n$ and Fermi momentum $k_F$)
as shown by Kadanoff and Baym \cite{Baym61,Baym62}.
It has been established, however, that the GW approximation fails to reproduce some key results
for the two-particle correlation functions and does not properly describe even the plasmon
properties, in contrast to RPA. This drawback has been clearly demonstrated for a
homogeneous electron gas (jellium model) by Holm and von Barth in Ref.~\cite{Holm98}.
Moreover, incorrect screening properties are expected to have a feedback on
single-particle spectra of real materials for which the GW approximation sometimes fails to
account for the observed value of the absolute band-gap \cite{Schone98}.

In this Letter we present a simple strategy to restore the physical two-particle correlation properties within the conventional GW approximation. Our trick can be applied at every order of the skeleton expansion, and does not produce any systematic bias in the infinite-order limit for convergent series. 
It could therefore in the future be used within a Diagrammatic Monte Carlo approach \cite{diagmc1,diagmc2,diagmc3}. 
We focus here on the jellium model, describing Coulomb-interacting electrons moving against a positively charged uniform background.
We found that previously published results for the ground state energy per particle $E/N$ obtained with the standard GW approach were in strong (well outside of reported error bounds) disagreement with each other,
see Refs.~\cite{Holm98,Holm2,Gonzalez,Yan}. The correct results were plotted only in Ref.~\cite{Yan}. We put an end to this detrimental situation and provide accurate values for the ground state energy, the quasiparticle $Z$-factor,
and the effective mass renormalization $m_*/m$ (where $m$ is the bare electron mass)
at the Fermi level.
%
\begin{figure}
\includegraphics[scale=0.38,width=0.99\columnwidth]{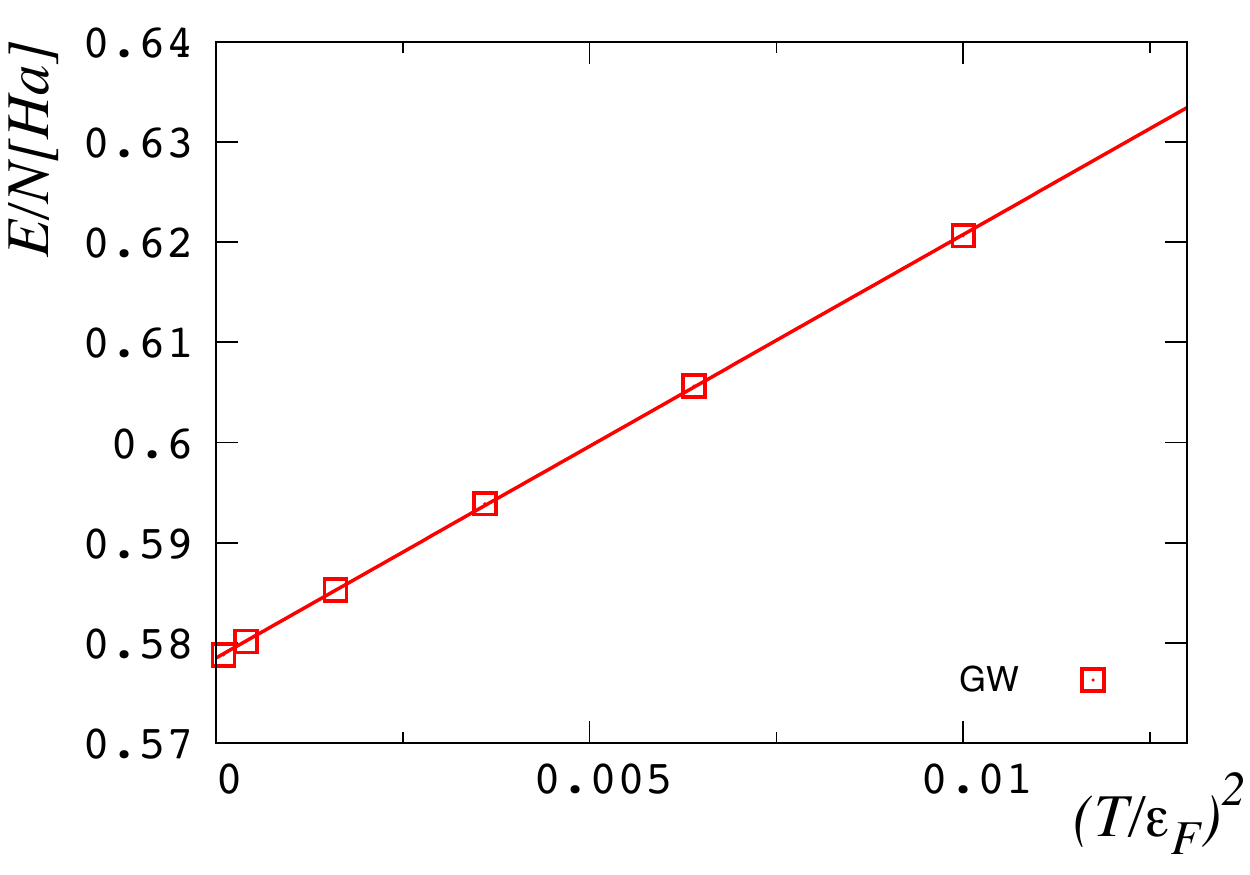}
\caption{Energy per electron (in Hartrees) as a function of $(T/\epsilon_F)^2$ revealing the Fermi-liquid behavior. The solid line is a linear fit giving a ground-state energy
of $E/N = 0.5783(2)$ Ha.
}
\label{fig:E}
\end{figure}

{\it Formalism.} Let us start by briefly reviewing the GW approximation. It is based on the lowest-order
skeleton diagrams for the irreducible self-energy $\Sigma_{\sigma}$
($\sigma$ is the spin index) and the irreducible polarization $\Pi$.
In the position-imaginary time $(r, \tau)$-representation it reads:
\begin{eqnarray}
 \Sigma_{\sigma}(r,\tau) & = &   -  G_{\sigma}(r, \tau) W(r, -\tau) \; ,\\
 \Pi(r, \tau) & = & \sum_{\sigma} G_{\sigma}(r, \tau) G_{\sigma}(r, -\tau) \; ,
 \label{GW}
\end{eqnarray}
where $G_{\sigma}$ is the one-body Green's function
and $W$ the effective screened interaction. These are self-consistently defined
through solutions of the Dyson equations in the momentum-Matsubara frequency
$(k, \omega_n)$-representation:
\begin{eqnarray}
G_{\sigma}(k, \omega_n)^{-1} & = &  G^0_{\sigma}(k, \omega_n)^{-1} - \Sigma_{\sigma}(k,\omega_n) \; , \\
W(k, \omega_n)^{-1} & = &  V(k)^{-1} - \Pi(k, \omega_n) \; ,
 \label{Dyson}
\end{eqnarray}
where $V(k) = 4 \pi e^2/ k^2$ is the bare Coulomb interaction
and $G^0_{\sigma}$ is the free one-body Green's function.
Knowing the one-body Green's function $G$ is sufficient for obtaining
the system's energy, as well as quasiparticle properties such as $m_*$ and $Z$, see Refs.~\cite{Fetter,Mahan}. We performed all calculations at finite temperatures
well below the Fermi energy $\epsilon_F$. For ground-state properties we extrapolated
results to zero temperature using the Fermi-liquid behavior.
In Fig.~\ref{fig:E} we show a typical plot for energy at $r_s = 1$, with standard definition
of $r_s$ as the ratio of the typical inter-particle spacing and the Bohr radius.

{\it Dielectric responce.} The work by Holm and von Barth \cite{Holm98} has established that the GW approximation
is not suitable for reliable analysis of two-particle correlation functions.
More precisely, it was found that the spectral function $S(k,\omega)$
of the irreducible polarization has incorrect behavior at frequencies $\omega > kv_F$, where
$v_F$ is the Fermi velocity; as a consequence, the real part of the dielectric function
$\epsilon (k, \omega ) =1-(4\pi e^2/k^2)\Pi(k,\omega )$
at small momenta $k\ll k_F$ has its zero shifted away from the plasmon frequency
$\omega_p = \sqrt{4\pi n e^2/m}$ to completely unphysical values, see Fig.~3 in Ref.~\cite{Holm98}.

Our results agree with this key observation: we also find that at $k\ll k_F$ and
$\omega_n \gg kv_F$ the irreducible polarization is {\it orders of magnitude}
larger than the expected values dictated by the plasmon mode, $ \Pi(k,\omega_n ) \approx -nk^2/m\omega_n^2$. This unphysical behavior can be traced back to the fact that the
GW approximation does not respect the particle number conservation law, which implies that
at zero momentum  $ \Pi(k=0,\omega_n ) \propto \delta_{n,0}$, or, identically,
$\Pi(k=0,\tau ) = {\rm const}$. [For an arbitrary interaction potential $\Pi $ is related
to the density-density correlation function $\chi$,
as $\Pi(k,\omega_n) = -\chi(k,\omega_n) /\left(1-V(k) \chi(k,\omega_n) \right)$, while $\chi (k=0, \tau) = {\rm const}$ or $\chi (k=0, \omega_n \ne 0) \equiv 0$.]
Instead, one finds that $ \Pi(k=0,\omega_n)$ has
significant amplitudes at finite frequencies, and, correspondingly,
$ \Pi(k,\omega_n\ne 0 )$ is not approaching zero when $k\to 0$.
This also causes significant problems for the proper technical implementation
of the GW approach in Coulomb systems because $ (4\pi e^2 /k^2)\Pi(k,\omega_n)$
tends to diverge as small momenta and forces one to consider extremely large frequencies
in the calculation of the screened interaction $W$.

Since all problems originate from the violation of the particle conservation law,
we propose a simple strategy to enforce the physical behavior of $\Pi(k,\omega_n)$.
All one has to do is to perform a transformation
\begin{equation}
\Pi(k,\omega_n) \to \Pi(k,\omega_n) -  \Pi(0,\omega_n) + \Pi(0,0) \delta_{n,0} \;, 
\label{trick}
\end{equation}
before calculating the dielectric response from the GW solution.
In other words, one has to subtract the spurious frequency dependence at $k=0$.
Note that this transformation is compatible with higher-order diagrammatics and
we suggest that it should be implemented within the fully self-consistent skeleton
schemes whenever one has to iterate properties of the $W$-function. Indeed, in the
large-order expansion limit the correction term is supposed to vanish as
$\Pi(k=0,\omega_n)$ converges to the correct physical behavior $\propto \delta_{n,0}$.
%
\begin{figure}[htb]
\includegraphics[scale=0.38,width=0.99\columnwidth]{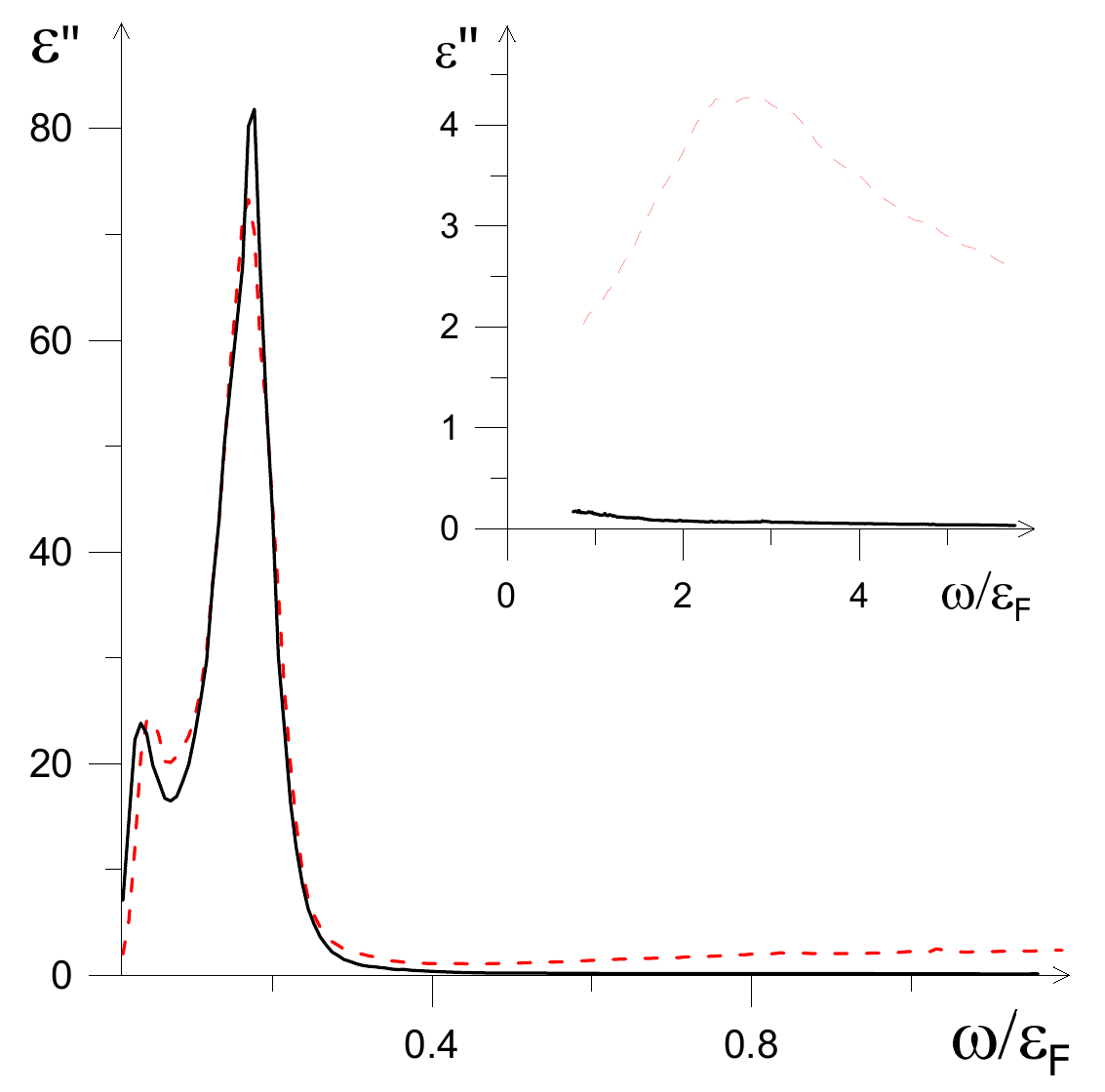}
\caption{ Color online:
Imaginary part of the dielectric function within the GW approximation at $r_s=1$,
$k/k_F=0.1$, and $T/\epsilon_F=0.02$. Red dashed curve is the original GW result, and
the solid black line is the corrected GW spectrum. The crucial difference at frequencies
$\omega > kv_F$ is clearly seen in the inset.
}
\label{fig:2}
\end{figure}
%
\begin{table*}
\caption{\label{tab:table3}
Minus the ground-state exchange-correlation energy per particle $-E_{XC}$ (in Hartree),
the quasi-particle residue $Z$, and the effective mass renormalization $m_*/m$
and  at the Fermi level for the unpolarized 3D homogeneous electron gas}
\begin{ruledtabular}
\begin{tabular}{cccccc}
 $r_s$     & $1$         & $2$         & $4$         & $5$           & $10$         \\ \hline
 $-E_{XC}$ & $0.5267(2)$ & $0.2789(1)$ & $0.1488(1)$ & $0.1216(1)$   & $0.06498(2)$ \\
 $Z$       & $0.899(1)$  & $0.842(1)$  & $0.769(2)$  & $0.743(2)$    & $0.658(2)$   \\
$m_*/m$   & $0.944(2)$  & $0.931(2)$  & $0.913(2)$  & $0.906(2)$    & $0.875(2)$  \\
\end{tabular}
\end{ruledtabular}
\label{tab:GW}
\end{table*}
\begin{figure}[htb]
\includegraphics[scale=0.38,width=0.99\columnwidth]{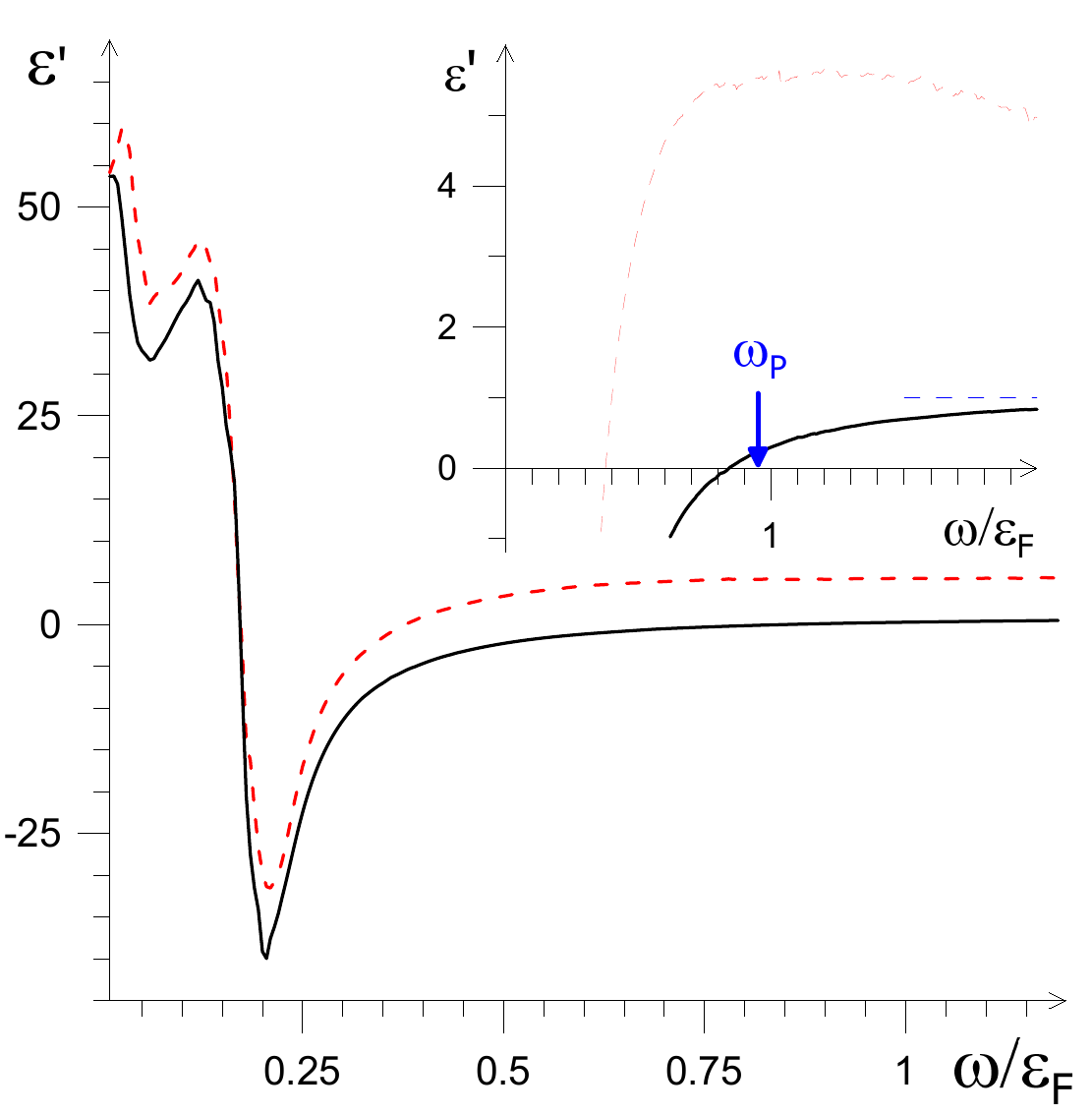}
\caption{Color online:
Real part of the dielectric function within the GW approximation at $r_s=1$,
$k/k_F=0.1$, and $T/\epsilon_F=0.02$. The original GW result (red dashed curve)
completely misses the plasmon zero, and predicts unrealistically large response
at frequencies above $\epsilon_F$. The corrected result (solid black line) crosses zero
within $10 \%$ of $\omega_p$ and saturates to unity at $\omega > \epsilon_F$.
}
\label{fig:3}
\end{figure}
%

In Figs.~\ref{fig:2} and~\ref{fig:3} we show how our protocol works in practice by considering
the case of $r_s=1$ at low temperature $T/\epsilon_F=0.02$ and small momentum $k/k_F=0.1$.
First, we performed analytic continuation of the imaginary frequency data for
$\epsilon (k, \omega_n)$ using  a hybrid of stochastic optimization \cite{SOM,Julich}
and consistent constraints \cite{CC} methods to get $\epsilon ''(k, \omega)$, and then
obtained the real part $\epsilon '(k, \omega)$ from the Kramers-Kronig relation.
The improvement in terms of eliminating the unphysical behavior is dramatic.
After the transformation, the high-frequency tail of $\epsilon ''(k, \omega)$
gets suppressed by nearly two orders of magnitude. 
As a result, the real part of the dielectric function now has its zero at
$\omega_p^{(GW)} \approx 0.89(1) \omega_p$
and is approaching unity from below at $\omega \gg \epsilon_F$.
[In order to have $\omega_p^{(GW)}$ to coincide with $\omega_p$ precisely,
one would need to divide $\Pi$ by $Z^2$, mimicking the effect of vertex
corrections.]
Everything about the original GW data at frequencies
$\omega > kv_F$ is completely unsatisfactory.

{\it Ground-state properties.} The GW technique, self-consistently solving the above set of Eqs.~(\ref{GW})-(\ref{Dyson}),
was implemented in the past for jellium at zero temperature in Refs.~\cite{Gonzalez, Holm98, Holm2}
and at finite temperature in Ref.~\cite{Yan}. It was concluded \cite{Holm2,Gonzalez}
that the method produces ground-state energies that agree with diffusion Monte Carlo
results \cite{Ceperley} at the sub-percent level. Apparently, this conclusion was based
on incorrect data. We find that our exchange-correlation energies differ from those of
Refs.~\cite{Gonzalez, Holm98, Holm2} by an amount bigger than the
difference between the GW and various other approximations, for instance GW$^{(0)}$.
To ensure correctness of our results, we developed two absolutely independent
codes that did not share a single common idea about grids and cutoffs for storing
and processing the data, Fourier transforms, and energy evaluation.
Moreover,
 the second code was implemented for the Yukawa potential and final results were recovered by extrapolating the Yukawa screening length to zero.
After requesting data used for plots in Ref.~\cite{Yan} we concluded that this was the only article reporting correct results for energy~\cite{Lode}.

For benchmark purposes we report here the ground-state exchange-correlation energy,
the quasi-particle $Z$-factor, and the effective mass renormalization in Table~\ref{tab:GW}.
Error bounds were estimated from variations induced by changing momentum-time grids,
cutoffs, and extrapolation procedures to the zero-temperature limit.
All results in the table were obtained for the {\it standard} GW formulation; i.e.,
the transformation procedure (\ref{trick}) was {\it not} applied when solving Eqs.~(\ref{GW})-(\ref{Dyson}).

{\it Conclusions.} We have proposed a simple strategy to drastically improve key properties
of the two-particle correlation functions within the GW approximation and applied it to
the jellium model. The very same trick can be applied to other models and materials
science systems, and can be used in the Diagrammatic Monte Carlo approach that considers
higher-order vertex corrections. We also report benchmark values of key Fermi liquid parameters
for jellium within the {\it standard} GW approximation.

{\it Acknowledgements.}
We thank Xin-Zhong Yan for sending us the $GW$ data presented in Ref.~\cite{Yan}.
This work was supported by
the Simons Collaboration on the Many Electron Problem, the National Science Foundation under
the grant PHY-1314735, and the MURI Program ``New Quantum Phases of Matter" from AFOSR.
A.S.M. was supported by ImPACT Program of Council for Science, Technology and Innovation
(Cabinet office, Government of Japan). N.P. acknowledges support from the Stiftelsen Olle Engkvist Byggm\"{a}stare Foundation, the Swedish Research Council grant 642-2013-7837
and the ERC grant Thermodynamix.



\begin{thebibliography}{99}
\bibitem{Lode} After this work was completed our values for the ground-state energy were confirmed by yet another (fourth) implementation by L. Pollet.
\end{thebibliography}
\end{document}